\documentclass[twocolumn,english]{article}
\usepackage[T1]{fontenc}
\usepackage[latin9]{inputenc}
\usepackage{geometry}
\usepackage{fancyhdr}
\usepackage{amsbsy}

\makeatletter
\newcommand{\lyxaddress}[1]{
\par {\raggedright #1
\vspace{1.4em}
\noindent\par}
}

\@ifundefined{date}{}{\date{}}
\makeatother

\usepackage{babel}
\begin{document}
\renewcommand\abstractname{\begin{flushleft}\textbf{\ \ \ \ \ Abstract}\end{flushleft}}
\twocolumn[\begin{@twocolumnfalse}

\vspace{-15mm}
\begin{center}
\small \bf{
International Symposium on Earth Science and Technology 2014
}
\vspace{-3mm}

\hrulefill
\vspace{-5mm}
\end{center}

\title{\bf A formulation for dissolution in inhomogeneous temperature field}

\author{Hiroki FUKAGAWA\textsuperscript{{*}}\textsuperscript{} and Takeshi
TSUJI\textsuperscript{$\dagger$}}

\maketitle

\lyxaddress{\begin{center}
International Institute for Carbon-Neutral Energy Research (I2CNER),
Kyushu University, Fukuoka 819-0395, Japan\linebreak{}
$^{*}$fukagawa.hiroki.609@m.kyushu-u.ac.jp\linebreak{}
$^{\dagger}$tsuji@i2cner.kyushu-u.ac.jp
\par\end{center}}
\begin{abstract}
\emph{We propose equations governing the dissolution in inhomogeneous
temperature field in terms of the variational principle. The derived
equations clarify that the interface energy between solute and solvent
has a significant effect on the process of the dissolution. The interface
energy restrains the dissolution, and the moving interface involves
the heat of dissolution.}\bigskip{}

\end{abstract}
\end{@twocolumnfalse}]
%\renewcommand\thefootnote{*}
%\footnotetext[1]{fukagawa.hiroki.609@m.kyushu-u.ac.jp}
%\renewcommand\thefootnote{$\dagger$}
%\footnotetext[2]{tsuji@i2cner.kyushu-u.ac.jp}
\thispagestyle{empty}
\pagestyle{empty}
%\onecolumn
%\twocolumn
\pagestyle{fancy}
\lhead{}
\chead{\footnotesize {\bf{CINEST 14-31\ \ \ }}}
\rhead{}
\cfoot{}

\section*{Introduction}

Dissolution is an important process and frequently discussed in industry
as well as in science. For example, the dissolution of supercritical
${\rm CO_{2}}$ in interstitial water is one of the most crucial research
topics in ${\rm CO_{2}}$ capture and storage (CCS), which stores
the ${\rm CO_{2}}$ into the deep underground in a geological rock
formation. The ${\rm CO_{2}}$ can be trapped in the micro-pore space
as droplets surrounded by water. At this small scale, the contribution
of the interface energy to the total energy is consequential, and
thus comes into play. Various phase field models based on free energies
are often used to study the dynamics on the assumption of constant
temperature and no heat transfer\cite{JDvanderWaals,cahn1958free}.
There can be phenomena involved in the inhomogeneous temperature and
the heat transfer. The diffusion flux can be induced by a temperature
gradient, which is known as the Soret effect or thermal diffusion\cite{PhysRev.37.405}.
The heat transfer during dissolution across the interface is also
considered important for the dynamics of the gas ${\rm CO_{2}}$ at
near the critical point because of its very large thermal conductivity\cite{guildner1958thermal}.
In previous works, heuristic methods have been proposed to combine
the thermodynamics with those phase field models above. In this study,
we propose a completely different method based on the variational
principle to derive the governing equations for the dissolution in
the inhomogeneous temperature field.

\section*{The variational principle}

The dynamics of a fluid can be divided into the kinetic part and thermodynamics
part. The kinetic part of the dynamics for the fluid is characterized
by the conservation laws for mass, energy, momentum, and angular momentum.
On the other hand, the thermodynamics is described by the equation
of the entropy in the form as
\begin{equation}
\frac{\partial}{\partial t}(\rho s)=\Theta-\nabla\cdot\boldsymbol{J},\label{eq:entropylaw}
\end{equation}
where $\rho$ is total mass density, $s$ is specific entropy, $\boldsymbol{J}$~is
entropy flux, and $\Theta>0$ is a dissipative function describing
entropy production rate per time. The equation (\ref{eq:entropylaw})
plays an important role in connecting the kinematics and thermodynamics.
In terms of the variational principle, we define the Lagrangian density
as the kinetic energy density minus the internal energy density, and
the action as the integral over space and time. The realized dynamics
minimizes the action under some constraints\cite{Fukagawa01052012,Fukagawa01092010,fukagawa2012,hyon2010energetic,liu2009introduction,2014arXiv1407.1035J}.
With the aid of (\ref{eq:entropylaw}), this principle enables us
to formulate the dynamics of the fluid even if it has complicated
constraints. Noether's theorem states that each of the conservation
laws is associated with each corresponding symmetry. For example,
the conservation laws for energy, momentum, and angular momentum are
related to the translation symmetries in time and space, and rotational
symmetry, respectively. Thus to satisfy these conservation laws, (\ref{eq:entropylaw})
has to be consist with these symmetries. If we don't know the exact
form of the dissipative function $\Theta>0$, we can fix it by considering
the symmetries. On the other hand, the entropy flux $\boldsymbol{J}$
is determined to erase surface terms without fixing boundary conditions
appearing in the variational calculus. Our method is very simple.
We just give the Lagrangian by the kinetic energy minus the internal
energy. The exact form of (\ref{eq:entropylaw}) is obtained by the
method above.

\section*{The two-component fluid\label{sec:Two-component-Fluid}}

We consider a two-component fluid composed of two substances: solute
and solvent. The conservation law of the total mass $\rho$ is given
by
\begin{equation}
\frac{\partial}{\partial t}\rho+\nabla\cdot(\rho\boldsymbol{v})=0,\label{eq:conservation law of rho}
\end{equation}
where $\boldsymbol{v}$ is mass average velocity. We introduce the
set of three scalers $\boldsymbol{A}=(A_{1},A_{2},A_{3})$ denoting
the initial position of the fluid particle at $(t,\boldsymbol{x})$.
By the definition, the material derivative $D_{t}\equiv\frac{\partial}{\partial t}+\boldsymbol{v}\cdot\nabla\ $
of $A_{i}$ is zero,
\begin{equation}
D_{t}A_{i}=0.\label{eq:Clebsch}
\end{equation}
In the variational calculus, we use $\boldsymbol{A}$ to describe
the path lines of the fluid particles, and fix the value of $\boldsymbol{A}$
at the boundary. Let $\psi$ be the mass fraction of the solute. The
mass conservation law of the solute is
\begin{equation}
\rho D_{t}\psi+\nabla\cdot\boldsymbol{j}=0,\label{eq:diffusion}
\end{equation}
where $\boldsymbol{j}$ is the diffusion flux of the solute. The diffusion
flux $\boldsymbol{j}$ describes the relative motion of the solute
and the solvent. Let $\boldsymbol{a}$ be the amount of the solute
flowing through the unit interface orthogonal to the direction of
$\boldsymbol{j}$, i.e.,
\begin{equation}
D_{t}\boldsymbol{a}-\boldsymbol{j}=0.\label{eq:diffusion-1}
\end{equation}
We also fix the value of $\boldsymbol{a}$ at the boundary. The main
purpose is to obtain the equations for $\boldsymbol{v}$ and $\boldsymbol{j}$
from the variational principle. We define the specific bulk internal
energy $\epsilon$ as the function of $\rho$, $\psi$, and $s$.
Here, $s$ is the specific entropy of the two-component fluid. Thus
we have 
\begin{equation}
d\epsilon=-Pd\rho^{-1}+\mu d\psi+Tds,
\end{equation}
in the thermodynamics. Pressure $P$ and temperature $T$ are defined
as $P\equiv\rho^{2}\left(\partial\epsilon/\partial\rho\right)_{s,\psi}$
and $T\equiv\left(\partial\epsilon/\partial s\right)_{\rho,\psi}$,
where the subscripts $_{s}$, $_{\rho}$ and $_{\psi}$ indicate variables
fixed in the respective partial differentiations. The coefficient
$\mu\equiv(\partial\epsilon/\partial\psi)_{s,\psi}$ is an appropriately
defined chemical potential of mixture, $\mu=\mu_{{\rm solute}}/\mu_{{\rm solute}}-\mu_{{\rm solvent}}/\mu_{{\rm solvent}}$,
where $\mu_{{\rm solute}}$ and $\mu_{{\rm solvent}}$ are the chemical
potentials of the two substances, and $m_{{\rm solute}}$ and $m_{{\rm solvent}}$
are the masses of the two kinds of the particles as in $\S$58 of
Ref.\,\cite{landau1959fm}. We write $E$ for the interface energy
density given as the function of $\rho,\psi$, and $\nabla\psi$,
and assume that $E$ is isotropic, i.e.,
\begin{equation}
\frac{\partial E}{\partial\partial_{i}\psi}\partial_{j}\psi=\frac{\partial E}{\partial\partial_{j}\psi}\partial_{i}\psi.\label{eq:iso-1}
\end{equation}
The internal energy density is the sum of the bulk energy density
and the interface energy density,
\begin{equation}
\rho\epsilon+E.\label{eq:newene}
\end{equation}
On the other hand, the total kinetic energy density is the sum of
the kinetic energy densities of the each fluid, and it is rewritten
into
\begin{equation}
\frac{1}{2}\rho\boldsymbol{v}^{2}+\frac{1}{2\rho}\left(\frac{1}{\psi}+\frac{1}{1-\psi}\right)\boldsymbol{j}^{2}.\label{eq:KE}
\end{equation}
The Lagrangian density ${\cal L}$ is given by subtracting (\ref{eq:newene})
from (\ref{eq:KE}),
\begin{equation}
{\cal L}\equiv\rho\frac{1}{2}\boldsymbol{v}^{2}+\frac{1}{2\rho}\left(\frac{1}{\psi}+\frac{1}{1-\psi}\right)\boldsymbol{j}^{2}-\left(\rho\epsilon+E\right).\label{eq:Le-1-1}
\end{equation}

Next, let us discuss the thermodynamics. Considering the translation
symmetries in time and space, which are respectively associated with
the conservation laws of energy and momentum, the equation of the
entropy (\ref{eq:entropylaw}) is given in the form of
\begin{equation}
\rho D_{t}s=\frac{\left(\sigma_{ij}\partial_{i}\nu_{i}-\nabla\cdot\boldsymbol{J}_{q}+\boldsymbol{\nu}\cdot D_{t}\boldsymbol{a}\right)}{T}-\nabla\cdot\boldsymbol{J}_{s}.\label{eq:deltaS-2-1}
\end{equation}
Here $\sigma$ and $\boldsymbol{\nu}$ are coefficients, and $\boldsymbol{J}_{q}$
is heat flux. Note that $v_{j}$ is the function of $\partial A_{i}/\partial t$
and $\partial_{j}A_{i}$ from (\ref{eq:Clebsch}). We determine  $\boldsymbol{J}_{s}$
as
\begin{equation}
\boldsymbol{J}_{s}=\frac{1}{T}\frac{\partial E}{\partial\nabla\psi}D_{t}\psi\label{eq:Js-1}
\end{equation}
to erase the surface term with respect to $\psi$ appearing in the
variational calculus of the Lagrangian (\ref{eq:Le-1-1}). Here $\partial E/\partial\nabla\psi$
takes large absolute value at the interface, and $D_{t}\psi$ expresses
the moving of the interface. Thus (\ref{eq:Js-1}) shows that entropy
flux occurs with accompanying the moving interface, which is related
to the heat of dissolution\cite{israelachvili2011intermolecular}.
The coefficient $\sigma$ is a symmetric tensor because of the rotational
symmetry corresponding to the conservation law of angular momentum.
We can rewrite (\ref{eq:deltaS-2-1}) in the form of (\ref{eq:entropylaw}).
Then $\Theta$ and $\boldsymbol{J}$ are respectively given by
\begin{eqnarray}
\Theta & \!\!\!\!=\!\!\!\! & \frac{1}{T}\left(\sigma_{ij}e_{ij}+\boldsymbol{\nu}\cdot\boldsymbol{j}\right)+\boldsymbol{J}_{q}\cdot\nabla\left(\frac{1}{T}\right),\label{eq:theta}\\
\boldsymbol{J} & \!\!\!\!=\!\!\!\! & \rho s\boldsymbol{v}+\frac{\boldsymbol{J}_{q}}{T}+\boldsymbol{J}_{s}.\label{eq:entropyflux}
\end{eqnarray}
Here $e_{ij}\equiv(\partial_{i}v_{j}+\partial_{j}v_{i})/2$ is the
strain rate tensor. We determine $\sigma$, $\nu$, and $\nabla T$
to make (\ref{eq:theta}) positive because of the second law of thermodynamics.
In the low degree approximation, (\ref{eq:theta}) is given by the
quadratic form of $e_{ij}$, $\boldsymbol{j}$, and $\nabla T$. If
we assume that $\sigma_{ij}$ depends on only $e_{ij}$ and is isotropic,
we have
\begin{equation}
\sigma_{ij}=2ae_{ij}+(b-2a/3)\delta_{ij}e_{kk},\label{eq:sigma}
\end{equation}
where $a$ and $b$ are the coefficients of shear and bulk viscosities,
respectively. If the both of $a$ and $b$ are positive, $\sigma_{ij}e_{ij}$
is also positive. Without loss of generality, we have
\begin{eqnarray}
\boldsymbol{\nu} & \!\!\!\!=\!\!\!\! & \xi\boldsymbol{j}+\eta\nabla T,\label{eq:nu}\\
\boldsymbol{J_{q}} & \!\!\!\!=\!\!\!\! & -\eta T\boldsymbol{j}-\kappa\nabla T,\label{eq:HeatFlux}
\end{eqnarray}
where $\xi$ is the coefficient of friction for the diffusion flux
$\boldsymbol{j}$, and $\kappa$ is the coefficient of thermal conductivity.
The coefficient $\eta$ in (\ref{eq:nu}) expresses the Soret effect
describing the flow of the solute induced by a temperature gradient.
On the other hand, the coefficient $\eta$ in (\ref{eq:HeatFlux})
shows the Dufour effect describing the energy flux due to the diffusion
flux $\boldsymbol{j}$ occurring. The both of $\eta$ in (\ref{eq:nu})
and (\ref{eq:HeatFlux}) expresses coupled effects of irreversible
processes. The coefficients $\xi$, $\eta$, and $\kappa$ are determined
to make (\ref{eq:theta}) positive\cite{PhysRev.37.405}.

The action is given by the integral of (\ref{eq:Le-1-1}) over the
considered time and space. By solving the stationary condition of
the action subject to (\ref{eq:conservation law of rho}), (\ref{eq:Clebsch}),
(\ref{eq:diffusion}), (\ref{eq:diffusion-1}), and (\ref{eq:deltaS-2-1}),
we obtain the equations of motion for the mass average velocity $\boldsymbol{v}$,
and the diffusion flux $\boldsymbol{j}$. The former is
\begin{equation}
\frac{\partial}{\partial t}\left(\rho v_{i}\right)+\partial_{j}\left(\rho v_{i}v_{j}+\Pi_{ij}+\sigma{}_{ij}\right)=0,\label{eq:momentum-2}
\end{equation}
where we use (\ref{eq:iso-1}), and write $\Pi_{ij}$ for
\begin{equation}
\Pi_{ij}=\left(P+\rho\frac{\partial E}{\partial\rho}-E\right)\delta_{ij}+\frac{\partial E}{\partial\partial_{i}\psi}\partial_{j}\psi.
\end{equation}
The latter is
\begin{eqnarray}
\!\!\!\!\!\!\!\!\!\!\!\!\!\! & \!\!\!\! & \!\! D_{t}\left\{ \frac{1}{\rho}\left(\frac{1}{\psi}+\frac{1}{1-\psi}\right)\boldsymbol{j}\right\} \nonumber \\
\!\!\!\!\!\!\!\!\!\!\!\!\!\! & \!\!=\!\! & \!\!-\nabla\left\{ \mu^{*}+\frac{1}{2\rho^{2}}\left(\frac{1}{\psi^{2}}-\frac{1}{(1-\psi)^{2}}\right)\boldsymbol{j}^{2}\right\} -\boldsymbol{\nu},\label{eq:DiffusionFlux-1}
\end{eqnarray}
where $\mu^{*}$ is the generalized chemical potential defined as
\begin{equation}
\mu^{*}\equiv\mu+\frac{1}{\rho}\frac{\partial E}{\partial\psi}-\frac{T}{\rho}\partial_{k}\left(\frac{1}{T}\frac{\partial E}{\partial\partial_{k}\psi}\right).\label{eq:GeneraizedMu}
\end{equation}
If the diffusion flux $\boldsymbol{j}$ is static and small, we have
\begin{equation}
\boldsymbol{j}=-\frac{1}{\xi}\nabla\mu^{*}-\frac{\eta}{\xi}\nabla T,\label{eq:j}
\end{equation}
from (\ref{eq:nu}) and (\ref{eq:DiffusionFlux-1}). The equation
(\ref{eq:j}) shows that the diffusion flux $\boldsymbol{j}$ occurs
in response to the gradients of the generalized chemical potential
$\mu^{*}$ and the temperature $T$. The third term in the right-hand
side in (\ref{eq:GeneraizedMu}) shows that the interface energy prevents
the dissolution of the solute, when the temperature $T$ is low.

\section*{Summary and Discussion}

We propose a new theoretical method based on the variational principle
for the two-component fluid in inhomogeneous temperature field. In
this proposed method, we combine the kinematics and thermodynamics
by using (\ref{eq:entropylaw}) in the variational calculus. In this
way, we obtain all the equations describing the whole dynamics of
the two-component fluid. We clarify that the interface energy plays
the important role in thermodynamics and dissolution as shown in (\ref{eq:Js-1})
and (\ref{eq:GeneraizedMu}), respectively. Previous theories based
on a free energy\cite{JDvanderWaals,cahn1958free} assume a constant
temperature and no heat flux in these theories, and cannot derive
the entropy flux (\ref{eq:Js-1}) and the generalized chemical potential
(\ref{eq:GeneraizedMu}). Our proposed method can be applied to various
more complicated fluids, and yields the governing equations consistent
with the conservation laws and thermodynamics\cite{fukagawa2014variational}.
What are required in our theory are the kinetic and the internal energy
densities. The exact form of (\ref{eq:entropylaw}) is determined
to satisfy symmetries and the second law of thermodynamics, and to
erase surface terms without fixing boundary conditions appearing in
the variational calculus. The equations of motion are derived from
the variational principle with the aid of (\ref{eq:entropylaw}).

\section*{Acknowledgments}

We gratefully acknowledge the support of the International Institute
for Carbon-Neutral Energy Research (WPI-I2CNER), sponsored by the
World Premier International Research Center Initiative (WPI), MEXT,
Japan, and appreciate the support of the SATREPS project by JICA-JST.

\bibliographystyle{unsrt}
\bibliography{ref}

\end{document}